\documentclass[aps,prl,twocolumn]{revtex4}
\usepackage[]{graphicx}
\usepackage{mathrsfs}
\usepackage{amssymb,amsmath}

\begin{document}
\title{Formation of optimal-order necklace modes in one-dimensional random photonic superlattices}

\author{Mher Ghulinyan}
\email[]{ghulinyan@itc.it}

\affiliation{Micro-Technology Lab., ITC-irst, via Sommarive 18,
I-38050 Povo (Trento), Italy}

\begin{abstract}
We study the appearance of resonantly coupled optical modes, optical
necklaces, in Anderson localized one-dimensional random
superlattices through numerical calculations of the accumulated
phase. The evolution of the optimal necklace order $m^{*}$ shows a
gradual shift towards higher orders with increasing the sample size.
We derive an empirical formula that predicts $m^{*}$ and discuss the
situation when in a sample length $L$ the number of degenerate in
energy resonances exceeds the optimal one. We show how the
\emph{extra} resonances are pushed out to the miniband edges of the
necklace, thus reducing the order of the latter by multiples of two.
\end{abstract}

\maketitle

Wave interference phenomena play a crucial role in transport
properties of various physical systems from periodic
\cite{soukoulis} to disordered \cite{Sheng,tigg}. Among these,
originally studied for electronic systems, Anderson localization
seems the more intriguing one \cite{Anderson}. It predicts a phase
transition form metallic-like conductivity to an insulating regime,
when the transport can come to halt with increasing the randomness.
On the other hand, in a one-dimensional (1D) disordered system, such
a transition happens when the sample size $L$ extends beyond the
so-called \emph{localization length} $\xi$. In such conditions, the
wavefunctions become localized within an extension $\xi$ and decay
exponentially with distance $l$. Being essentially an interference
phenomenon, Anderson localization has been studied as for
electromagnetic and acoustic waves \cite{localization}, as well as
for degenerate atomic gases \cite{degas}. Very recently, Anderson
localization of optical waves in the microwave regime has been
demonstrated in experiments on 1D random multilayer dielectric
stacks \cite{necklace1,necklace2}.

Initially, it has been widely accepted that the conductivity
(transmittivity) of a disordered chain is mainly supported by states
which are closer situated to the sample center \cite{note1}. Later,
this was questioned, since for long enough specimen, the states in
the center possess significantly reduced probability to support a
two-step hopping transport through these states. In late 80's,
\emph{Pendry} \cite{p87} and \emph{Tartakovskii et al.} \cite{tart},
independently, argued that the conductivity in such systems should
be dominated by so-called necklaces -- few homogeneously distributed
through the sample states, degenerate in energy, and coupled
resonantly to delocalize and extend through the chain. The number
$m$ of resonant states forming a necklace, was calculated to scale
as $L^{1/2}$, while the probability of their occurrence was shown to
drop as $exp(-L^{1/2})$, thus predicting them to be increasingly
improbable events in long samples \cite{p87}. Therefore, for a
certain sample length, a trade-off between the expected number $m$
and their occurrence probability would determine the optimal order
of the necklace.

In this Letter we study the appearance of optical necklaces in
finite 1D random superlattices through numerical calculations of the
accumulated phase and follow the evolution of the optimal necklace
order when increasing the sample size. We suggest an empirical
formula which predicts the optimal necklace order $m^{*}$ and
discuss the situation when in a sample length $L$ the number of
degenerate in energy resonances exceeds $m^{*}$. We show that the
\emph{extra} resonances, which are spaced by a distance less than
the optimal one for a certain sample length, are coupled strongly
enough to split and be pushed out to the miniband edges of the
necklace, thus reducing the order of the latter by multiples of two.

We studied random binary multilayer stacks composed by $\mathcal{N}$
layers of $A$ and $B$-type dielectric materials (refractive indices
chosen to be $n_A=1.3$ and $n_B=2.1$). Positionally random sequences
were generated by giving equal weights to each type of layer. Two
types of random multilayer stacks were investigated. In one type,
the physical thicknesses $d_A$ and $d_B$ of the constituent layers
were generated randomly through the sequence, thus a complete
randomness was achieved. For the other type, $d_A$ and $d_B$ were
chosen to be resonant at some wavelength $\lambda_0$ (quarter-wave
layers), therefore, the condition $n_A d_A=n_B d_B=\lambda_0/4$ was
considered. The quarter-wave condition makes the system not
perfectly random and, in particular, maximizes the constructive
interference effects at the resonant wavelength.

\begin{figure} [t!]
\includegraphics[width=8.5cm]{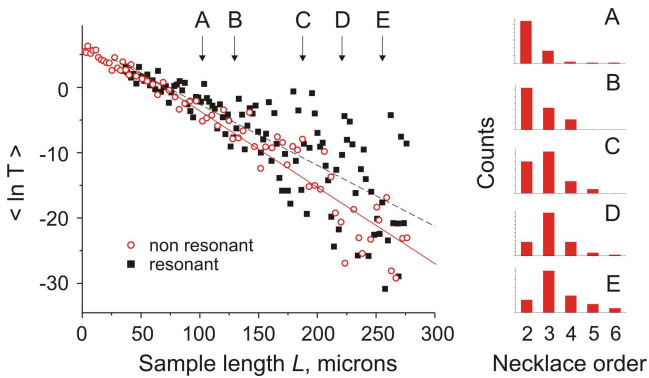}\\
\caption{ (Color online) The logarithm of integrated transmission
versus sample length for two cases of non-resonant ($\bigcirc$) and
resonant ($\blacksquare$) layers. The solid and dashed lines are
linear fits to the data for non-resonant and resonant layers cases,
correspondingly. The fits give localization lengths of
$\sim8.5~\mu$m and $\sim10.7~\mu$m respectively. The histograms on
the right show the counts of the order of observed necklaces for
five different thicknesses of samples with resonant $\lambda/4$
layers.}\label{neck}
\end{figure}

In general, in a 1D dielectric multilayer stack, the phase suffers a
$\pi$ shift each time it meets a resonance. Therefore, the phase
measurement is a valid tool to isolate the spectrum singularities in
either periodic \cite{prbmatteo} or random systems \cite{necklace2}.
This way, in Ref.~\cite{necklace2}, localized states and, more
importantly, optical necklaces have been identified through
interferometric measurements on Anderson localized 1D random
superlattices. When $m>2$ degenerate in energy states couple weakly
to form a necklace \cite{note2}, the accumulated phase increases
smoothly through the transmission band built up by these latter,
summing up to $m\pi$. On the contrary, when strong coupling occurs,
the phase suffers single $\pi$ jumps through well defined spectrally
separate peaks.

Hence, through this study, the order of the necklaces has been
assigned to the multiples of $\pi$ in the smoothly changing phase
shift through the transmittance band. Firstly, random structures,
with non-resonant layer thicknesses, were studied and their
transmission spectra were calculated through a standard transfer
matrix method \cite{pAdv}. In Fig.~\ref{neck} the logarithm of the integrated
transmission over a wide spectral range ($1\div2~\mu$m), averaged
over 200 realizations per point, is plotted versus the sample
thickness (open circles) \cite{note3}. We observe, firstly, that
$<\ln T>$ decreases as $L^{-1}$, as it is expected in the diffusive
regime \cite{Sheng}, where the states are still extended. Over about
30 $\mu$m a smooth transition from extended to localized regime
occurs and the sample transmission drops linearly with further
increase in the sample thickness. The localization length of
$\xi\approx8.5~\mu$m is obtained from the slope of $<\ln
T>=-\frac{L}{\xi}$.

In order to study the properties of optical necklaces, one has to go
through many realizations of disorder, because of their nature to be
extremely rare events. For this, we decided to look for necklaces in
binary multilayer stacks with resonant quarter-wavelength layers. We
were motivated by the fact that, due to the resonant layer
condition, these provide much higher probability to possess many
resonances at $\lambda_0$ wavelength.

As it is seen from Fig.~\ref{neck}, the transmission of such
realized samples (full squares) shows a clear deviation from the one
of fully random system. Namely, the higher-on-average transmission
in this case is supported by much frequent occurrences of optical
necklaces at $\lambda_0$ and becomes more pronounced for thicker
samples. We have verified the order of the necklaces in a small
frequency window around the resonant wavelength for five different
sample lengths. For this, 300 realizations per each chosen thickness
have been studied and the necklaces have been identified through the
calculated phase. The right panels in Fig.~\ref{neck} report the
counts of different order necklaces. We observe that at smaller
thicknesses (panel A, $L\approx103~\mu$m) the major number of
necklaces is of second order and fewer third order ones occur. With
increasing sample size, the number of second order necklaces
diminishes and the third order ones firstly start to dominate at
$L\approx 184~\mu$m, obtaining a maximum at $L\approx220~\mu$m. With
a further increase in sample size (panel E, $L\approx253~\mu$m)
higher orders shoulder starts to rise in the necklace counts
histogram, indicating that the samples become enough thick to fit
more than three degenerate resonances.

\begin{figure} [b!]
\includegraphics[width=6.5 cm]{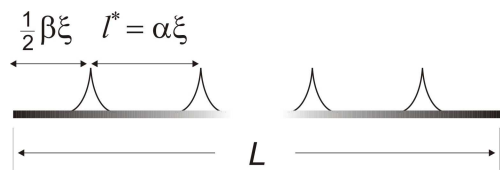}\\
\caption{ A simplified model of formation of an optimal-order
necklace through the sample length $L$.}\label{emp}
\end{figure}

The order of the necklace, which fits optimally inside the sample
length $L$ to have the highest and most compact transmission band,
can be empirically calculated as
\begin{equation}\label{floor}
    m^{*} =\lfloor \frac{L-\beta\xi}{\alpha\xi} \rfloor+1,
\end{equation}
where $\lfloor~\rfloor$ stand for the \emph{floor}-function and
denote the greatest integer part of the expression inside. Here, the
parameter $\alpha$ counts the distance (in the units of $\xi$)
between two neighboring resonances efficiently coupling to form a
necklace, while $\beta$ considers the coupling of first and last
resonances to the environment (see Fig.~\ref{emp}). In reality, the
step function of Eq.~\ref{floor} considers the order of the most
frequent necklace from the histogram of all observed orders. In our
case, a reasonable fit to the numerical results was found with
$\alpha=7$ and $\beta=2.2$.

\begin{figure} [t!]
\includegraphics[width=7.8 cm]{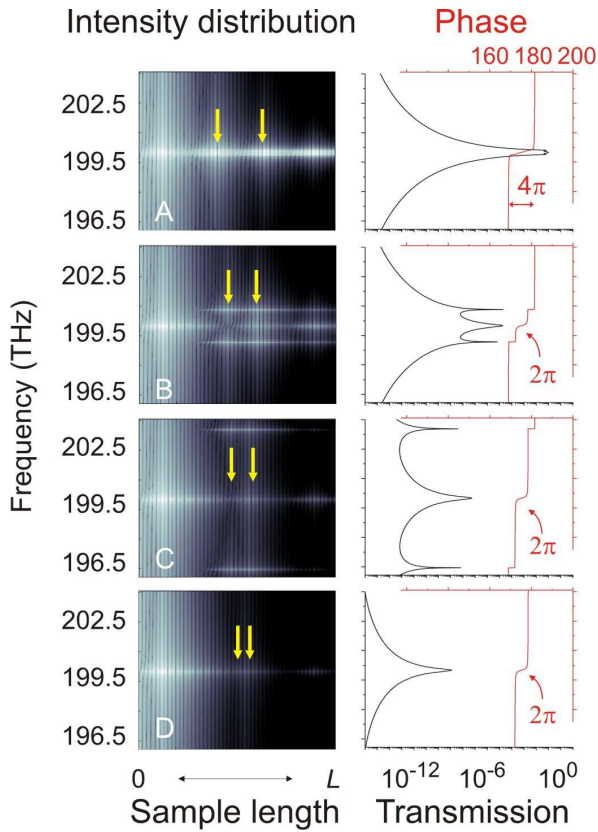}\\
\caption{ (Color online) The light intensity distribution inside a
periodic superlattice of four coupled cavities (left panels) and the
corresponding transmission and phase spectra (right panels). All the
spectra are plotted in a the narrow frequency range around
$\nu_0=c/\lambda_0=200$~THz. (a) homogeneously distributed cavities
(an artificial $m$=4 necklace) form the highest transmission band
through which the phase accumulates smoothly a $4\pi$ increase,
(b)-(d) moving the two central cavities closer to each other, the
light transmission around $\lambda_0$ attenuates by orders of
magnitude and the smooth phase jump counts only $2\pi$, reducing the
necklace order. }\label{close}
\end{figure}

The power law scaling of the order of necklaces in disordered
systems and the very low probability of the higher orders occurrence
predicts a trade-off scenario, which should reduce the number of
actually observed ones \cite{p87}. In the following, we address this
issue to understand how this reduction happens and reveal
interesting physical picture of the phenomenon.

Let us consider the case of an optimal $m$th order necklace
stretching through sample length $L$. The most compact and, in the
meanwhile, high-transmission band formed by this should occur when
$m$ resonances, degenerate in energy, are distributed homogeneously
through the sample to favor similar coupling between neighboring
states. This situation is an ideal one and is very improbable for
higher order necklaces. Since the resonances occur randomly in the
depth of the sample, it becomes more probable that some of them will
appear at a distance closer than the optimal separation
$l^{*}=\alpha\xi$ between two neighboring cavities. When this
happens, the spatially close resonances couple strongly and a larger
mode repulsion pushes them farther from the miniband center.

We examine the dynamics of such a situation on an example of a
periodic photonic structure, where for simplicity four resonant
cavities are coupled through Bragg sequences in such a way that a
high transmission band of an artificially built necklace is formed.
The choice of a periodic structure allows us to manipulate the
spatial positions of various cavities, which is an impossible task
for random systems. Figure~\ref{close} shows the light intensity
distribution through the photonic structure (left panels) and the
corresponding transmission spectrum together with the phase (right
panels). In Fig.~\ref{close}A the ideal situation is presented, when
the resonances are distributed homogeneously through the sample. A
high transmission miniband is formed (left panel, black line)
through which the phase (red) suffers a smooth increase by $4\pi$.

Next, we move spatially the two central cavities closer to each
other. The intensity distribution plot in Fig.~\ref{close}B shows
the effect of stronger coupling between these two resonances: two
well defined resonances (symmetric and anti-symmetric modes) rise
out of the miniband core due to the mode repulsion and the channel
intensity decreases. The transmission of the miniband peak drops by
four orders of magnitude, and the smooth change in the phase now
sums up to $2\pi$. If we move these cavities closer
(Fig.~\ref{close}C,D), the peak transmission drops down to $\sim
10^{-8}$. Note that when excluding the two central cavities (while
maintaining the same sample length) the transmission drops further
down by almost three orders of magnitude to have a peak transmission
of $\sim 10^{-11}$.

The intensity map in Fig.~\ref{close}D contains an interesting
feature, to which we would like to draw a special attention. We
observe three maxima instead of two through the original
transmission miniband, while the phase counts always a $2\pi$ jump.
We stress, that the central intensity maximum is not due to a
resonance present at $\lambda_0$, but is built by the overlap
integral of the transmission lines of the symmetric and
anti-symmetric states. Therefore, it is correct to consider the
necklace at $\lambda_0$ to be of second rather than of third order
one.

\begin{figure} [t!]
\includegraphics[width=8.5 cm]{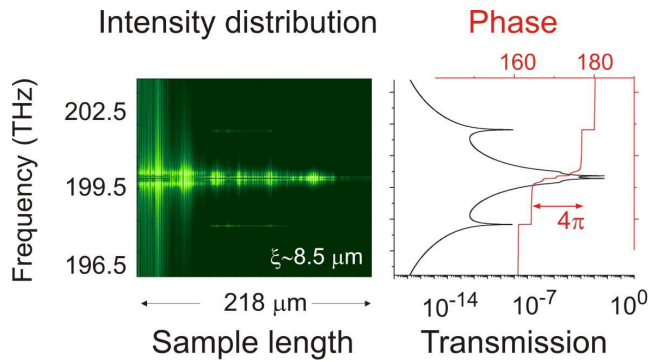}\\
\caption{ (Color online) An example of a necklace formed in a
$218~\mu$m-thick random system. The intensity distribution map
(left) shows six bright clumps, while the phase (right) counts four
$\pi$'s, indicating the formation of a fourth order necklace.
}\label{r950}
\end{figure}

In order to catch such a situation in a random system, we have
performed a number of realizations of a $950$ layer sample
($218~\mu$m-thick). In Fig.~\ref{r950} we show such an event. In the
right panel six well-resolved intensity maxima are observed, while
the smooth change of the phase sums up to $4\pi$. The intensity
distribution map, and, therefore, the electric field profile at the
miniband center frequency shows many bright and other less intense
clumps, but the real number of resonances, directly involved in the
light transport through the miniband is less and can be exactly
counted through the phase variation.

Thus, when the number of randomly built degenerate cavities exceeds
the optimal order of the necklace for a certain sample length, the
extra cavities couple strongly to be pushed out to the edges of the
miniband, reducing the order of the necklace state and its peak
transmission. We stress that this reduction is always a multiple of
two, independently on the fact whether even or odd number of extra
cavities are coupled strongly. This is because a strong coupling of
even or odd number of cavities results in a dip or a peak in the
spectral line at the resonant wavelength. Therefore, at $\lambda_0$,
only multiples of $2\pi$ are filtered out from the phase jump.
Nevertheless, the extra cavities play an important role in the
formation of necklaces, since their non-zero overlap reduces the
reflectivity of inter-cavity regions and contributes positively to
link the resonant tunneling transport through the necklace miniband.

Along this, in a fully random system, another interesting situation
can happen. Suppose, that next to a necklace band centered at
$\lambda_0$ (call $N^0$), for simplicity, two other degenerate
states, resonant at some wavelength $\lambda_0+\Delta$ (or
$\lambda_0-\Delta$), are formed. If these are coupled strongly
enough, it can happen that a fortuitous mode splitting will push
either the symmetric or the antisymmetric mode to fit into the band
width of $N^0$. In the case, when the original necklace at
$\lambda_0$ is week, the new state can positively contribute to
enhance the necklace transmission, thus increasing the order of
$N^0$ by one. This is to say that the phase shift will count only
one more resonance, but still the intensity map will show extra
nods, since the new resonance will appear with a double clump. We
note, that in a random sample with \emph{resonant} layers, such a
situation cannot occur: the transmission spectrum in this case is
always symmetric against $c/\lambda_0$, therefore, the described
contribution to the necklace $N^0$ will occur from both sides
($\lambda_0\pm\Delta$) and consequently will result in a back
repulsion of degenerate modes out of the necklace band.

To conclude, we address through this study some peculiarities of the
formation of optical necklaces, resonantly coupled degenerate modes,
in one-dimensional random multilayer systems. We stress the
importance of using the phase-jump method to reveal the exact order
of necklaces, in contrast to counting intensity or electric field
clumps through the necklace. We show how the extra resonances are
filtered out from the necklace when in a sample length $L$ the
number of degenerate in energy resonances exceeds the optimal one.

We acknowledge helpful discussions with Lorenzo Pavesi, Diederik
Wiersma, Georg Pucker and Pierluigi Bellutti.


\begin{thebibliography}{99}
\bibitem{note1}
This was motivated by the fact that the probability for a particle
to be transmitted by hopping in and out of the sample is the highest
for the states  originating in the chain center.
\bibitem{note2}
We refer to a necklace, when the coupling between constituent states
is weak enough to result in a mode repulsion, smaller than the full
width at half maximum of individual modes.
\bibitem{note3}
As discussed in Ref.~\cite{necklace1}, assuming ergodicity, one can
consider the wavelength average of the transmission spectrum, to
avoid ensemble averages, since for the latter many realizations are
required.
\end{thebibliography}
\end{document}